\documentclass[12pt]{article}

\usepackage[letterpaper,hmargin=1in,vmargin=1in]{geometry}

\usepackage{graphicx,epstopdf,amsmath,amsfonts,amssymb}

\def\be{\begin{equation}}
\def\ee{\end{equation}}
\def\ba{\begin{eqnarray}}
\def\ea{\end{eqnarray}}
\def\ge{\mathrel{\raise.3ex\hbox{$>$\kern-.75em\lower1ex\hbox{$\sim$}}}}
\def\la{\mathrel{\raise.3ex\hbox{$<$\kern-.75em\lower1ex\hbox{$\sim$}}}}

\def\simgt{\mathrel{\raise.3ex\hbox{$>$\kern-.75em\lower1ex\hbox{$\sim$}}}}
\def\simlt{\mathrel{\raise.3ex\hbox{$<$\kern-.75em\lower1ex\hbox{$\sim$}}}}

\newcommand{\bi}[1]{\bibitem{#1}}
\newcommand{\fr}[2]{\frac{#1}{#2}}

\newcommand{\nub}{\mbox{$\nu_b$}}

\newcommand{\nc}{\newcommand}

\nc{\gone}{\bar g_{\pi NN}^{(1)}}
\nc{\gzero}{\bar g_{\pi NN}^{(0)}}
\nc{\al}{\alpha}
\nc{\ga}{\gamma}
\nc{\de}{\delta}
\nc{\ep}{\epsilon}
\nc{\ze}{\zeta}
\nc{\et}{\eta}
\nc{\ka}{\kappa}
\nc{\rh}{\rho}
\nc{\si}{\sigma}
\nc{\ta}{\tau}
\nc{\up}{\upsilon}
\nc{\ph}{\phi}
\nc{\ch}{\chi}
\nc{\ps}{\psi}
\nc{\om}{\omega}
\nc{\Ga}{\Gamma}
\nc{\De}{\Delta}
\nc{\La}{\Lambda}
\nc{\Si}{\Sigma}
\nc{\Up}{\Upsilon}
\nc{\Ph}{\Phi}
\nc{\Ps}{\Psi}
\nc{\Om}{\Omega}
\nc{\ptl}{\partial}
\nc{\del}{\nabla}
\nc{\ov}{\overline}
\nc{\newcaption}[1]{\centerline{\parbox{15cm}{\caption{#1}}}}
\nc{\us}{U(1)$_S$}
\nc{\ub}{U(1)$_B$}
\nc{\co}{CoGeNT}
\nc{\ctw}{$^{12}$C}
\nc{\cth}{$^{13}$C}
\nc{\ctwm}{{\rm ^{12}C}}
\nc{\cthm}{{\rm ^{13}C}}
\nc{\neff}{${\cal  N}_{\rm eff}$}
\nc{\neffm}{{\cal  N}_{\rm eff}}
\nc{\bore}{$^{8}$B}

\def\beq{\begin{equation}}
\def\eeq{\end{equation}}
\def\bmat{\begin{displaymath}}
\def\emat{\end{displaymath}}
\def\bear{\begin{eqnarray}}
\def\eear{\end{eqnarray}}
\def\ba{\begin{eqnarray}}
\def\ea{\end{eqnarray}}
\def\bery{\begin{array}}
\def\ery{\end{array}}
\def\bit{\begin{itemize}}
\def\eit{\end{itemize}}
\def\ben{\begin{enumerate}}
\def\een{\end{enumerate}}
\def\btab{\begin{tabular}}
\def\etab{\end{tabular}}
\def\btbl{\begin{table}}
\def\etbl{\end{table}}
\def\bfig{\begin{figure}[htb]}
\def\efig{\end{figure}}
\def\bpic{\begin{picture}}
\def\epic{\end{picture}}

\def\ga{\mathrel{\raise.3ex\hbox{$>$\kern-.75em\lower1ex\hbox{$\sim$}}}}
\def\la{\mathrel{\raise.3ex\hbox{$<$\kern-.75em\lower1ex\hbox{$\sim$}}}}
\def\gappeq{\mathrel{\rlap {\raise.5ex\hbox{$>$}}
{\lower.5ex\hbox{$\sim$}}}}
\def\lappeq{\mathrel{\rlap{\raise.5ex\hbox{$<$}}
{\lower.5ex\hbox{$\sim$}}}}

\def\gyr{{\rm \, G\kern-0.125em yr}}
\def\mev{{\rm \, Me\kern-0.125em V}}
\def\gev{{\rm \, Ge\kern-0.125em V}}
\def\tev{{\rm \, Te\kern-0.125em V}}

\begin{document}

\begin{titlepage}

\setcounter{page}{1}

\vspace*{0.2in}

\begin{center}

\hspace*{-0.6cm}\parbox{17.5cm}{\Large \bf \begin{center}
Neutrino Physics with Dark Matter Experiments and \\
the Signature of New Baryonic Neutral Currents
\end{center}}

\vspace*{0.5cm}
\normalsize

\vspace*{0.5cm}
\normalsize

{\bf Maxim Pospelov$^{\,(a,b)}$}

\smallskip
\medskip

$^{\,(a)}${\it Perimeter Institute for Theoretical Physics, Waterloo,
ON, N2J 2W9, Canada}

$^{\,(b)}${\it Department of Physics and Astronomy, University of Victoria, \\
     Victoria, BC, V8P 1A1 Canada}

\smallskip
\end{center}
\vskip0.2in

\centerline{\large\bf Abstract}

New neutrino states \nub, sterile under the Standard Model interactions, 
can be coupled to baryons via the isoscalar vector currents that are much stronger than 
the Standard Model weak interactions. 
%because new baryonic interactions can be much stronger than SM weak interactions.  
If some fraction of solar neutrinos oscillate into \nub\ on their way to Earth, 
the coherently enhanced elastic \nub-nucleus 
scattering can generate a strong signal in the dark matter detectors. 
For the interaction strength a few hundred times stronger than the weak force, 
the elastic \nub-nucleus scattering via new 
baryonic currents may account for the existing anomalies in the 
direct detection dark matter experiments at low recoil. We point out that for
solar neutrino energies the baryon-current-induced inelastic scattering is suppressed,
so that the possible enhancement of new force  is not in 
conflict with signals at dedicated neutrino detectors.
We check this explicitly  by calculating  the \nub-induced deuteron breakup,
and the excitation of 4.4 MeV $\gamma$-line in $^{12}$C. Stronger-than-weak 
force coupled to baryonic current implies the existence of new abelian 
gauge group \ub\ with a relatively light gauge boson.

\vfil
\leftline{March 2011}
    
\end{titlepage}

\subsection*{1. Introduction}

Standard Model (SM) of particles and fields must be augmented 
to include neutrino mass physics and perhaps extended 
even further to account for the "missing mass" of the Universe, or cold dark matter (DM). 
During the last decade the underground experiments 
\cite{cdms,xe,DAMA,cogent,cresst,zep,edel,picasso} aimed at direct detection 
of DM in the form of weakly interacting massive particles (WIMPs) \cite{WIMPS} 
have made significant inroads into 
the WIMP-nucleon cross section vs WIMP mass parameter space. Since no  DM-induced nuclear recoil signal was found 
with the exception of two hints to be discussed below, 
they constrained many models of dark matter and ruled out some portion of the 
parameter space in the best motivated cases such as {\em e.g.} supersymmetric neutralino 
DM \cite{witten}, Higgs-portal singlet DM \cite{HiggsDM} etc. 

It has been argued by some authors  that although primarily designed to 
search for WIMPs, these experiments are in fact multi-purpose devices that 
can be also used for alternative signatures of other effects beyond Standard 
Model. In particular, using same instruments one can look for the 
absorption of keV-scale bosonic super-WIMPs \cite{sWIMPs}, 
search for the axion emission from the Sun \cite{aSun},
and also investigate some additional signatures of WIMP-atom scattering that exist in 
"non-minimal" WIMP models \cite{nonMin}. This paper extends this point 
further and opens a new direction: we show that neutrino physics beyond SM can also 
be probed with the dark matter experiments. 

Elastic scattering of neutrinos on nuclei \cite{Stodolsky} is enhanced by the coherence factor $N^2$, 
where $N$ is the number of neutrons. 
Straightforward analysis \cite{FM,nuDM} of the SM solar neutrino elastic scattering rates on nuclei 
used in DM experiments reveal
several basic points:\\
\begin{itemize}
\item  Despite the coherent enhancement, the scattering rates are way too small, 
leading to counting rates not exceeding 
$10^{-3}~{\rm kg}^{-1}{\rm day}^{-1}{\rm keV}^{-1}$. Such low rates do not introduce 
any $\nu$-background to WIMP searches at the current levels of sensitivity. 
\item  The nuclear recoil spectrum is usually very soft, $E_r\sim (E_\nu)^2/M_{\rm Nucl} ~
\sim {\rm few~KeV}$ or less.
\item Solar boron ($^8$B) neutrinos are the best candidates for producing an observable signal,
because of the compromise between the relatively large flux and the energy spectrum extending to 15 MeV.
\end{itemize} 
Of course, at this point the DM experiments typically target much harder recoil
and are far away from low counting rates induced by solar neutrinos. On the other hand, 
the last generation of dedicated solar neutrino experiments \cite{SuperK,SNO,Borexino} have been 
extremely successful in 
detecting solar neutrinos via charged current reactions (CC),  elastic scattering on electrons (ES), 
and  $Z$-exchange mediated (NC) deuteron break-up \cite{SNO}. 
It is the combination of all these three signals that led to a very credible resolution of 
the long-standing solar-neutrino deficit problem via the neutrino oscillation and the 
MSW mechanism \cite{osc,MSW}. 

However, it is easy to imagine that three active SM species $\nu_e,~\nu_\mu,~\nu_\tau$ with 
their (almost completely) established mass/mixing parameters may not be the last word in the 
neutrino story. In this paper we consider a model of "quasi-sterile" neutrino $\nu_b$ that 
has no charged currents with normal matter, no ES on electrons or other leptons, but has much 
enhanced NC with baryons (NCB).
We shall consider the strength of new NCB interaction to be much larger than Fermi constant, 
$G_B \sim (10^2-10^3)\times G_F$. Such interactions can be mediated by new vector 
bosons of the gauged baryon number \ub, and for that reason we call this new hypothetical neutrino 
state as the "baryonic" neutrino \nub. If any of the solar SM neutrino flavors
oscillate into $\nu_b$ within one astronomical unit, then the current DM experiments will
in principle be able to pick it up via the coherently enhanced NCB signal. Whether such strong NCB would lead to a 
measurable energy deposition in the standard neutrino experiments requires special investigation and 
is addressed in this paper. We find that although the inelastic NCB scattering is enhanced by 
a huge factor $G_B^2/G_F^2$, it is also suppressed by a tiny factor $E_\nu^4 R_N^4$, 
where $R_N$ is a nuclear radius-related parameter. The resulting rate for NCB processes in
neutrino detectors is then can be made comparable or smaller than the regular neutrino counting rates.

It is somewhat tempting to relate the proposed $\nu_b$ model with the recently reported 
anomalies/signals in the direct DM detection. For a long time of course the DAMA and its successor
DAMA/LIBRA experiments have been claiming \cite{DAMA} the annual modulation of the energy deposition 
in NaI crystals with the maximum in early June and minimum in December, which is consistent with the 
expected seasonal modulation of the WIMP-nucleus scattering rate. Then, last year the CoGeNT collaboration \cite{cogent} 
has reported
unexpected (in the null hypothesis) rise of their signal at recoil energies below $E_r=$1 keVee. Given the mass of 
Ge nuclei, and typcial quenching factors in germanium, 
it is plausible that the rise of \co\ signal at low KeVee can be produced 
by \nub\ resulting from oscillations of boron neutrinos, $\nu_{SM} \to \nu_b$,
and hypothesized enhancement of NCB can compensate for small neutrino scattering rates. 
It is also clear that mimicking DAMA signal with $\nu_b$ is also possible in a more restricted sense.
Of course the usual seasonal modulation of the neutrino flux due to the eccentricity 
of the Earth orbit will have a minimum in the early July and a maximum in early January. 
However, the neutrino oscillation phenomenon is   not monotonic in distance \cite{Kamland},
and if the oscillation length for $\nu_{SM} \to \nu_b$ is comparable to the Sun-Earth distance, the annual modulation phase of the 
\nub\ scattering signal can be reversed by $\pi$. We investigate this 
opportunity, and conclude that both \co\ and DAMA signal can be described with \nub-type models
(provided that DAMA data can tolerate a $\sim 1$ month phase shift). 
We further argue that if indeed this is the case, the model is very predictive, and 
there will be further ample opportunites for probing  $\nu_b$ both at DM and neutrino detectors,
as well as at more conventional particle physics experiments.  

This paper is organized as follows: in the next section we introduce the class of \nub\ models
and specify a parameter range that is the most perspective for the DM experiments. Section 3 
addresses NCB elastic and inelastic scattering, including calculations of 
\nub-induced activation of carbon and the deuteron breakup reactions. 
Section 4 studies the possibility of phase reversal in the annual modulation
signal. Our conclusions are reached in Section 5.

\subsection*{2. Baryonic neutrino and baryonic neutral currents}

The basic features of the model are as follows: we introduce a new gauge group 
\ub, and give all quark fields of the SM, $Q,~U, D$, the same charge under 
\ub, which we call $g_b/3$. We also introduce a new left-handed neutrino species $\nu_b$
that has a charge $g_l$ under this new group and no charge under any of the SM gauge groups. 
In the interest of anomaly cancellations it is also 
desirable to introduce a right-handed partner of $\nu_b$ with the same charge. Then the new group couples 
to the "vector-like" matter multiplets, and although SM+\ub\ will in general be anomalous, the anomalies 
can be cancelled at some heavy scales. Variants of this model may include some partial 
gauging of the SM lepton species under \ub. Neither the right-handed $\nu_b$ nor the details 
of anomaly cancellation will be important for this paper. 
Furthermore, we assume that \ub\ is spontaneously broken by the \ub-Higgs vacuum expectation value $\langle \phi_b \rangle$, 
and exactly how this happens will not be of direct consequence for us either. 
The relevant gauge part of the Lagrangian is then given by
\be
\label{start}
{\cal L} = - \fr14V_{\mu\nu}^2 + \fr12m_V^2V_\mu^2 + \bar \nu_{b} \gamma_\mu(i\partial_\mu + g_lV_\mu) ~\nu_{b}  
+\sum_{q} \bar q (iD_{SM}
\!\!\!\!\!\!\!\!\!\!\!/ \;\;\;\;\;
+\fr{1}{3}g_b \gamma_\mu V_\mu) q + {\cal L}_m.
\ee
First two terms in (\ref{start}) are the standard Maxwell-Proca terms for $V_\mu$, the sum extends over all 
quark types and flavors, and $D_{SM}$ is the SM covariant derivative that includes
gauge interactions appropriate for each quark species $q$. The mass part of the 
Lagrangian ${\cal L}_m$ besides the usual SM mass terms  should also account for 
neutrino masses, and generate mixing to a new state \nub. In this paper we are not going to consider vector 
exchange with virtualities beyond $O(10~{\rm MeV})$, and therefore it is convenient to 
switch from quarks to nucleons, 
\be
\label{isosc}
 \fr{1}{3}V_\mu g_b~\sum_{q} \bar q  \gamma_\mu q ~\to~ g_b V_\mu(\bar p \gamma_\mu p + \bar n \gamma_\mu n) + ...
 \ee
Ellipses stands for $O(m_N^{-1})$ terms associated with the $V_{\mu\nu} \bar N \sigma_{\mu\nu}N$ 
part of the form factor, which will be small for any process we consider in this paper. 
The coupling of $V_\mu$ to the isoscalar vector current of nucleons 
$J_\mu^{(0)}=\bar p \gamma_\mu p + \bar n \gamma_\mu n$ will have important implications 
for both the elastic and inelastic scattering of \nub\ on nuclei. The exchange by the \ub\ gauge boson
creates the NCB Lagrangian, 
\be
\label{exchange}
{\cal L}_{NCB} = \bar \nu_b \gamma_\mu \nu_b ~\fr{g_lg_b}{m_V^2 + \Box} ~J_\mu^{(0)},
\ee 
that in the limit of $m_V^2\gg Q^2$ is just a new contact dimension 6 operator with the 
effective coupling constant $G_B$:
\be
{\cal L}_{NCB} = G_B\times \bar \nu_b \gamma_\mu \nu_bJ_\mu^{(0)};~~
G_B = \fr{g_l g_b}{m_V^2}  \equiv {\cal N} \times \fr{10^{-5}}{\rm GeV^2}.
\ee
Here we have introduced an "enhancement" parameter ${\cal N}$ that quantifies how much stronger 
$G_B$ is compared to the weak-scale value of $10^{-5}{\rm GeV}^{-2}$. We note that stronger-than-weak 
interactions among four neutrino species were considered earlier in {\em e.g.} Ref. \cite{Khlopov0}. 
The use of baryonic force as a mediator between SM and dark matter was considered recently in \cite{Ko}. 

One may wonder if ${\cal N}$ as large as a 100 or a 1000 can be consistent with low energy data
on meson decays, such as $K\to \pi \bar \nu_b \nu_b$. It turns out that due to the conservation 
of the baryon current, the loop amplitude for the underlying $s\to d \bar \nu_b \nu_b$ decay is 
additionally suppressed by $G_F Q^2$, which compensates for all possible enhancements due to ${\cal N}$.
(In contrast, the quark axial vector analogue of (\ref{start}) will be strongly constrained to 
have ${\cal N} \la 1$.) Thus, from quark flavor perspective, the baryonic portal (\ref{start}) is 
one of the two "safe" portals (the other being the kinetic mixing with hypercharge \cite{Holdom}) 
that allow attaching stonger-than-weak interactions to the quark currents. We shall not pursue 
the meson decay constraints on the model any further in this paper, and turn our attention 
to the neutrino mass sector. 

The most natural way of having a UV-complete theory of neutrino masses is via the introduction
of right-handed neutrinos states $N_R$. We can use the same singlet right-handed neutrinos 
coupled to the Higgs-lepton bilinears $LH$ and Higgs$_b$--neutrino \nub\ bilinears $\nu_b \phi_b$
in a gauge-invariant way,
\be
{\cal L}_m = LH {\bf Y} N  +  \nu_{bL}\phi {\bf b} N +(h.c.) + \fr{1}{2} N^T{\bf M}_RN.
\ee
Here ${\bf M}_R$ and ${\bf Y}$ are the familiar 3$\times$3 right-handed neutrino mass matrix and Yukawa matrix, 
while ${\bf b} $ is the new Yukawa vector parametrizing the couplings of the left-handed part of $\nu_b$ to $N$. 
Integrating out $N$ states results in the low-energy  4$\times$4 neutrino  mass and 
mixing matrices, $M_{ij}$, where $i,j$ run over $e,\mu,\tau,b$ flavors. 
While of course a full four-state analysis can be done, we shall simplify 
our discussion by the following assumptions: 
\begin{enumerate}

\item The entries of $3\times 3$ submatrix $M_{\rm active,~active}$ will in general be somewhat larger than 
$M_{{\rm active},b}$ and $M_{b,b}$ components so that the mixing pattern can be addressed 
sequentially: first the mixing of the SM neutrinos and then the admixture of the $\nu_b$. 

\item     A tri-bimaximal ansatz will be taken for the 3$\times$3 mixing of the SM neutrino species
for simplification, although having $\theta_{13}=0$ is not crucial. 

\item We shall assume a preferential mixing of 
$\nu_b$ to $\nu_2$, with the relevant parameters that we call $\Delta m_b^2$ and $\theta_b$, 
so that true mass eignestates are $\nu_I = \cos\theta_b \nu_2 + \sin\theta_b\nu_b$; 
$\nu_{II} = -\sin\theta_b \nu_2 + \cos\theta_b\nu_b$.

\item The sign of $G_B$ will be chosen to ensure that the matter effects for $\nu_b$ 
will not lead to the matter-induced $\nu_{\rm active} \to \nu_b$ transitions. 

\end{enumerate} 

The combination of these assumptions forms the following (simplified) picture of neutrino oscillations: 
inside the Sun the neutrino oscillations occur largely between $\nu_e$ and 
$\nu_+ \equiv (\nu_\mu+\nu_\tau)/\sqrt{2}$, 
\be
\left(\begin{array}{c}
\nu_e\\\nu_+
\end{array}
\right) \simeq 
\left(\begin{array}{cc}
\sqrt{\fr{2}{3}}& \sqrt{\fr{1}{3}}\\-\sqrt{\fr{1}{3}}&\sqrt{\fr{2}{3}}
\end{array}
\right)
\left(\begin{array}{c}
\nu_1\\\nu_2
\end{array}
\right)
\ee
while the "$-$" combination and $\nu_b$ stay unexcited. We would need only the higher-end of the Boron neutrino 
spectrum where MSW effect dominates. 
Upon the neutrino exit from the dense region of the Sun, it represents 
an almost pure $\nu_2$ state,  $ \nu_2 = \sqrt{\fr{1}{3}}\nu_e + \sqrt{\fr{2}{3}}\nu_+$, with individual 
flavor probabilities 
\be
P_e({\rm Sun}) \simeq \fr{1}{3};~~ P_+({\rm Sun}) \simeq \fr{2}{3};~~ P_b({\rm Sun})=0.
\ee
Then vacuum oscillations start building a non-zero probability for \nub\ due to 
the $\nu_I$ and $\nu_{II}$ being the true mass eigenstates in vacuum. Upon the arrival to the Earth, the following 
 energy-dependent probabilities will approximate the neutrino flavor composition:
 \begin{eqnarray}
 P_b({\rm Earth}) \simeq  \sin^2(2\theta_b)\sin^2\left[\fr{ \Delta m^2_b L(t)}{4E }\right] \nonumber\\
 \label{P's}
 P_e({\rm Earth}) \simeq \fr{1}{3}\left( 1 - \sin^2(2\theta_b) \sin^2\left[\fr{ \Delta m^2_b L(t)}{4E }\right] \right)\\
 P_+({\rm Earth}) \simeq \fr{2}{3}\left( 1 -  \sin^2(2\theta_b)\sin^2\left[\fr{ \Delta m^2_b L(t)}{4E }\right] \right), \nonumber
 \end{eqnarray}
 where $L(t)$ is the Earth-Sun distance with a slight eccentricity modulation, 
 \be
 L(t) \simeq  L_0\left(1-\epsilon \cos\left[
\fr{2\pi(t-t_0)}{T} \right]\right); ~~ L_0 = 1.5\times 10^8{\rm km}; ~~ \epsilon \simeq 0.167;~~ t_0 \simeq 3~{\rm Jan} .
 \ee
The most interesting range for $\Delta m^2_b$ to consider is
\be
\label{hier} 
 10^{-10} ~{\rm eV}^2  \la   \Delta m^2_b ~\ll ~ \Delta m^2_{\rm Solar,~atm}.
\ee
A scale of $O(10^{-10}) ~{\rm eV}^2 $ is the so-called "just so" mass splitting that may introduce significant 
changes to the otherwise very predictable $\propto L^{-2}$ seasonal 
variations of the \nub\ flux at Earth's location. With $\Delta m^2_b$ being much smaller than 
$10^{-5}~{\rm eV}^2$ there is no danger of distorting KamLAND results \cite{Kamland} even for a 
relatively large angle $\theta_b$, although the matter effects for $\bar\nu_b$ could be significant.
We also find it convenient to introduce the energy parameter $E_0$ directly related to the mass splitting,
\be
E_0 = \fr{\Delta m_b^2 L_0}{4\pi} = 6.05~{\rm MeV} \times \fr{\Delta m_b^2}{10^{-10}~{\rm eV}^2},
\ee
that defines last zero of $P_b$ as the function of energy, $P_b(L=L_0,E=E_0) = 0$.
Since in all NCB rates $P_b$ will enter in the combination with $G_B^2$, it is 
also convenient to define
\be
{\cal  N}_{\rm eff}^2 = {\cal N}^2 \times \fr{1}{2}\times \sin^2(2\theta_b),
\ee
so that in the limit of large $E_0$ the oscillations average out and 
$P_bG_B^2 \to {\cal  N}_{\rm eff}^2 \times 10^{-10}~{\rm GeV}^{-4}$.

The pattern of masses and mixing considered here is not the most natural: 
we assume a pair of very degenerate $\nu_{I}$ and $\nu_{II}$ mass eigenstates 
replacing $\nu_2$ and \nub. Given that the mass of $\nu_2$, regardless of the hierarchy pattern,
is always in between $\nu_1$ and $\nu_3$, this will require some specific adjustments of 
the full 4$\times$4 mass matrix. The search for more natural realizations of 
$\nu_{SM} \to \nu_b$ oscillations with long oscillation length, including matter effects for 
a different sign of $G_B$, goes beyond the 
scope of this paper. The goal of the next two sections will be to find the sensitivity to \neff\ 
in various processes involving the elastic and inelastic scattering of \nub.

We would like to close this section with some model-building comments. 
A very intriguing question to ask is whether SM neutrinos would tolerate new large 
NCB. A conventional answer is "no", as the so-called non-standard neutrino 
interactions (NSI) with quarks and charged leptons were addressed in a number of papers \cite{NSI} 
and almost no room at $O(1)G_F$ level was found, let alone the much enhanced NCBs hypothesized in this paper. 
However, NSI studies \cite{NSI} with rare exceptions \cite{Nelson} assume that the scale of the 
mediation is comparable to the weak scale, and ignore the possibility of light 
vector bosons communicating between neutrinos and baryons. As a matter of counterexample one could consider 
a model with two new gauge groups, \ub\ and the other being a quantized lepton flavor,
{\em e.g.} $L_\mu$ or $L_\tau$. The connection between two new vector sectors is given by the 
kinetic mixing term $\eta V^{(1)}_{\mu\nu} V^{(2)}_{\mu\nu}$. Then additional effective interaction of 
a SM neutrino with the baryonic current is given by 
\be
{\cal L}_{\rm eff} \propto 
\bar \nu_{SM} \gamma_\alpha \nu_{SM} ~\fr{\eta g_lg_b \Box}{(m_{V1}^2 + \Box)(m_{V2}^2 + \Box)} ~J_\alpha^{(0)}
\label{howdoyoulikeit}
\ee
Such interaction gives no contribution to the forward scattering amplitude and thus is not affecting neutrino 
oscillation, and it is $1/Q^2$ suppressed in the large $Q^2$ regime, avoiding strong constraints from
deep-inelastic neutrino scattering. It is then clear that the 
choice of $m_{V1}, ~ m_{V2}$ in the MeV range may allow having (\ref{howdoyoulikeit}) at 
$Q^2 \sim (1-10)~ {\rm MeV}^2$ to be considerably stronger than the SM weak force\footnote{The author would like to acknowledge 
very stimulating discussions with B. Batell and I. Yavin on the possibility of the NSI enhancement.}. 
The interactions of type (\ref{howdoyoulikeit}) can lead to the 
detectable recoil signal from elastic scattering 
of solar SM neutrinos, along the same lines as the \nub-scattering idea advocated in this paper.
A possibility of modified SM neutrino interactions such as (\ref{howdoyoulikeit}) can be 
very effectively tested 
%down to 
%$G_F$ level 
using the proposed neutrino-nucleus elastic scattering detectors placed near the intense 
source of stopped pions \cite{CLEAR}.

\subsection*{3. Elastic and inelastic scattering of \nub}

Elastic scattering of $\nu_b$ on nuclei will create a recoil signal regulated by the 
strength of NCB, and the probability of oscillation (\ref{P's}). It can be picked up by the 
direct dark matter detection experiments with low recoil thresholds. 
Also, \nub\ neutrinos can deposit a significant amount of energy 
on the order of a few MeV by activating excited nuclear states or via extra neutrons 
created by nuclear breakup.  
The main finding of this section can be summarized as follows: the ratio of the 
elastic to inelastic cross sections in the interesting neutrino energy range $E_\nu \la 15$ MeV, 
is governed by the following relation:
\be
\fr{\sigma_{\nu_b-{\rm Nucl}}({\rm elastic})}{\sigma_{\nu_b-{\rm Nucl}}({\rm inelastic})}\sim \fr{A^2}{E_\nu^4R_N^4} \sim 10^8,
\label{ratio}
\ee
where we took $A\sim 100$, $R_N^{-1} \sim 100 $ MeV, and $E_\nu \sim 10$ MeV.
It is this huge ratio that makes small-scale experiments such as \cite{cogent} competitive in 
sensitivity to \nub\ with the large-scale neutrino detectors. 

\subsubsection*{3.1. Elastic scattering}

The differential cross section for the NCB elastic scattering of left-handed \nub\ on a nucleus of mass 
$M_N$ with $A$ nucleons is given by
\be
\fr{d\sigma_{\rm el}}{d(\cos\theta)}= \fr{E^2A^2g_b^2g_l^2(1+\cos\theta)}{4\pi(M_V^2 +{\bf q}^2)^2}
\simeq \fr{1}{4\pi}\times G_B^2E^2A^2(1+\cos\theta),
\label{elastics}
\ee
where the elastic scattering momentum transfer is $q = ({\bf q}^2)^{-1/2} = 2 E\sin(\theta/2)$ and cannot 
exceed twice the neutrino energy $E$. In the second relation we took $M_V \gg E$, which allows to shrink vector propagator. 
Using relations between the neutrino scattering angle $\theta $, nuclear recoil energy $E_r$,
and the minimum neutrino energy required to produce $E_r$-recoiling nucleus,
\be
E_r = \fr{E^2}{m_N}\times(1-\cos \theta);~ E^{\rm min} = \sqrt{\fr{E_r M_N}{2}},
\ee
we rewrite the elastic cross section (\ref{elastics}) in the following form
\be
\fr{d\sigma_{\rm el}}{dE_r}= \fr{1}{2\pi}\times G_B^2A^2m_N\left(1-\fr{(E^{\rm min})^2}{E^2}\right).
\label{NCBel}
\ee
One can readily see that the NCB cross section (\ref{NCBel}) is related to the SM elastic neutrino-nucleus 
cross section \cite{Stodolsky}  by 
$ G_B^2 A^2 \to G_F^2 (N/2)^2$ substitution, where $N$ is the number of neutrons
(with small corrections in $1-4\sin^2\theta_W$ parameter). For the momentum transfers and nuclei considered 
in this paper the form factor corrections are $<5\%$ percent and are ignored. 

Using cross section (\ref{NCBel}), the flux and the energy distribution
of \bore\ neutrinos \cite{Bahcall} ({\em hep} solar neutrinos provide a small correction),
we derive the counting rates as the function of interaction strength and the oscillation probability.
For a moment, we neglect small seasonal modulations and take the limit of $\epsilon\to 0$ . 
For the medium composed only of atoms 
with atomic number $A$, we approximate these rates by:
\begin{eqnarray}
\fr{dR}{d E_r} \simeq \fr{A^2m_N}{2\pi}\times \fr{1}{2}\sin^2(2\theta_b)G_B^2\Phi_{\rm ^8B}\times I(E_r,E_0)\nonumber\\
\simeq 85 ~\fr{\rm recoils}{\rm day\times kg\times KeV} \times 
\left( \fr{A}{70} \right)^3\times \fr{\neffm^2}{10^4}\times I(E_r,E_0).
\label{dR}
\end{eqnarray}
The input (total flavor) flux of \bore\ neutrinos is taken to be $\Phi_{^8{\rm B} } = 5.7\times 10^6{\rm cm}^{-2}{\rm s}^{-1}$
and $m_N \propto Am_p$.

The recoil integral $I(E_r,E_0)$ in eq. (\ref{dR}) is given by the convolution of the \bore\ energy 
distribution, the energy-dependent part  of oscillation probability, and the kinematic factor in the 
cross section reflecting neutrino helicity conservation:
\be
I(E_r,E_0) = \int_{E^{\rm min}(E_r)}^\infty dE ~
\left(1-\fr{(E^{\rm min})^2}{E^2}\right) \times f_{\rm ^8B}(E) \times 2\sin^2\left[\fr{ \pi E_0}{E }\right].
\label{IEr}
\ee
Here the distribution function is normalized as $\int_{{\rm all} ~ E} f_{^8{\rm B}}(E)dE = 1$. For the limit of large $E_0$ (fast oscillations), the last multiplier in (\ref{IEr}) becomes 1. If a detector threshold 
corresponds to recoil energies that require $E^{\rm min}$ to be above the end-point of \bore\ neutrino spectrum,
$I\equiv 0$ (apart from small corrections from {\em hep} and diffuse supenova neutrinos). It is the case for most of 
the existing WIMP detectors, but not for all of them.

In real detectors registering ionization such as \cite{DAMA,cogent} 
it is the electron equivalent of the 
energy release rather than the recoil energy that is detected. We take the relation between the 
two by following recent DM-related analyses \cite{Itay,Collar}
\begin{eqnarray}
{\rm Ge:}~~~E_r({\rm keVee}) \simeq 0.2\times (E_r({\rm keV}))^{1.12}\nonumber\\
{\rm Na ~in ~NaI:} ~~~ E_r({\rm keVee}) \approx 0.33\times E_r({\rm keV}).
\label{quenching} 
\end{eqnarray}
The second relation is far less precise than the first one \cite{Collar}.

The counting rates in germanium resulting from scattering of \nub\ created by the oscillations of 
\bore\ and {\em hep} solar neutrinos are presented in Fig. 1. 
%with  and
%$\Phi_{hep } \sim 9\times 10^3{\rm cm}^{-2}{\rm s}^{-1}$. 
%The {\em hep} neutrino flux 
%plays only a minor role, and its uncertainty is not important. 
We have taken three cases of mass splitting:
$E_0 \gg E_\nu^{\rm max}$, and $E_0 = 12, ~14$ MeV. 
The NCB rates are plotted for the value of $\neffm^2 = 10^4$. 
For this enhancement factor, the resulting counting rates are clearly within 
reach of current  generation of low-threshold germanium detectors ({\em i.e.} \co). 

\begin{figure}
%\centering
\includegraphics[width=0.9\textwidth]{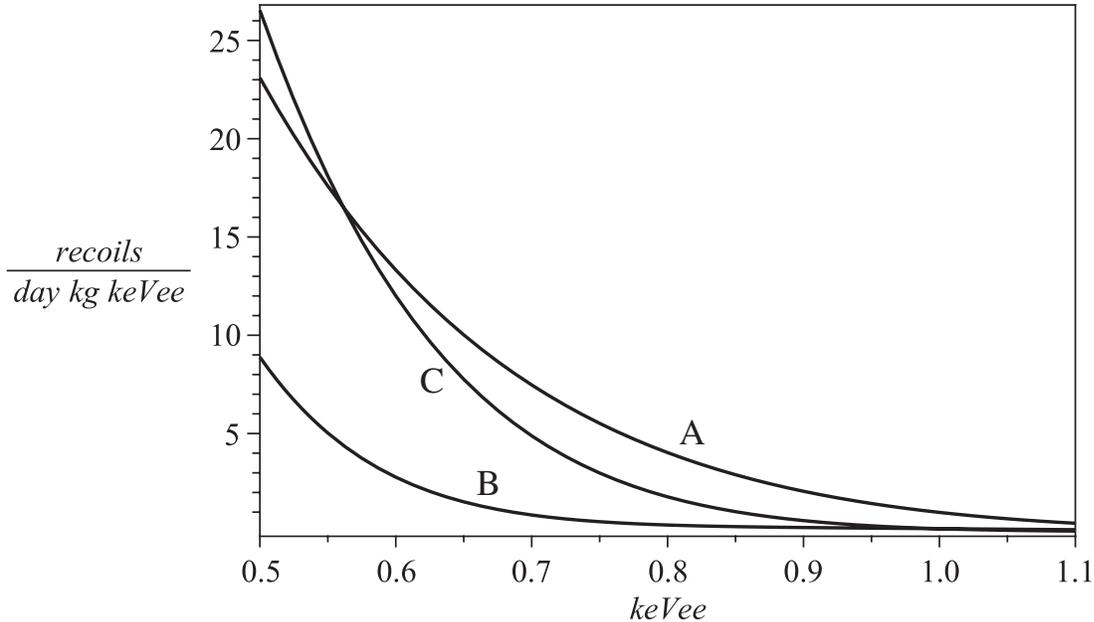} 
\caption{Expected recoil event rate in  Germanium in units of recoils/day/kg/keVee as the function of 
 $E_r$ in keVee. The NCB enhancement factor, $\neffm=100$. A, B and C lines correspond to 
$E_0=\infty$, $E_0=12$ MeV and $E_0 = 14$ MeV. }
\label{Ge}
\end{figure}

Inspection of Figure 1 shows that choice of different mass splitting that makes the oscillation 
length comparable to 1 A.U.   influences 
the shape of the spectrum. This is because the most important part of the 
spectrum for the recoil in excess of 0.5 keVee is above neutrino energies of 10 MeV, where the \bore\ neutrino 
spectrum is already sharply falling. The neutrino oscillations with $E_0$ close to 12 MeV will lead to the 
suppression of higher $E_r$ and to the steep rise of the signal at lower $E_r$. The sharp 
end of the neutrino spectrum prevents other Ge experiments with higher threshold like CDMS \cite{cdms} to 
probe the NCB scattering in the regime of large recoil where \cite{cdms} has strong sensitivity. 
The signal from the recoil due to \nub\ neutrinos is very similar in morphology to that of sub-10 GeV scale WIMPs. 
This is because a typical momentum transfer in a heavy nucleus - light WIMP collision is 
$q \sim m_{\rm wimp} v \sim 10$ MeV, which is about the same for \bore\ neutrino scattering. 
There is one kinematic difference though: the back-scattering of WIMPs that produces hardest 
recoil is kinematically allowed, while for neutrinos it is forbidden by helicity 
conservation. This additionally limits the capabilities of high-threshold experiments to detect 
\nub\ neutrinos in comparison with light WIMPs.

Is it possible to use \nub\ as another speculative explanation of \co\ results \cite{cogent}? 
The overall event rate can indeed be reproduced well with $\neffm \sim 10^2 - {\rm few}\times 10^2$,
depending on $E_0$. For the large $E_0$ parameter, the enhancement factor of $\neffm = 10^2$
seems sufficient: it gives 7 recoils/day/keVee at $E_r = 0.7$ keVee, which is about the same as 
the experimental data suggest after accounting for the efficiency \cite{cogent,Collar}. 
The shape of the predicted signal is also similar to the counting rate profiles observed by 
CDMS at the Stanford Underground facility \cite{SUF}.
Fitting the exact spectral shape of excess events 
at \co\ falls outside the scope of our current investigation. We should also note that the 
expected total counting rate for the material used in CRESST detectors \cite{cresst}
due to the neutrino-oxygen scattering is given by 
\be
R_{\rm O ~in~ CaWO_4}(E_r> 10~{\rm keV}) \simeq 0.2\times \fr{\rm recoils}{\rm day\times kg} \times \fr{\neffm^2}{10^4},
\ee
which is well within their detection capabilities for $\neffm \sim 100$.
Other methods in development that use liqud helium as a detecting medium with 
a potentially very low energy threshold \cite{he} also look  promising for detecting \nub-induced recoil.
It is also important that the choice of very low-mass target such as $^4$He will allow discriminating
between $\ga 5$ GeV WIMPs and \nub's: the effective recoil energy goes down 
at $M_N<M_{\rm WIMP}$, while it becomes larger for \nub\ scattering. 

\subsubsection*{3.2. Inelastic scattering} 

Unlike light WIMPs that can cary significant momentum but very little energy, \nub\ can easily lead to 
an MeV-scale energy deposition. 
Here we turn our attention to the NCB inelastic processes and will adress the following issues: 
the NCB deuteron breakup, and the NCB excitation of the first $2^+$ resonance in $^{12}$C resulting in 4.4 MeV $\gamma$ line:
\begin{eqnarray}
\label{breakup}
d+\nu_b &\to& \nu_b+n+p\\
^{12}{\rm C} +\nu_b &\to& \nu_b +^{12}{\rm C}^*(4.44~{\rm MeV}) \to \nu_b + ^{12}{\rm C}+ \gamma 
\label{excitation}
\end{eqnarray}
The main scientific question to answer is whether the enhanced values of $G_B^2P_b$ can be consistent with the 
constraints provided by SNO on "extra neutrons" from (\ref{breakup}) and by Borexino and other liquid scintillator detectors 
on "extra gammas" from (\ref{excitation}). There are of course other processes that one has to consider in a more 
comprehensive study, including the excitation of oxygen, the breakup of $^{13}$C to $^{12}$C$+n$ etc, but they will all 
follow the scaling in Eq. (\ref{ratio}). The earlier studies of the nuclear excitations due to the 
differenty type of neutrino couplings can be found in \cite{Khlopov1}. There are also elastic channels of energy deposition 
via $\nu_b+p\to \nu_b +p$ \cite{JohnPetr}, but the proton recoil from the scattering of \bore\ neutrinos 
would fall below the detector thresholds.

To understand the origin of ratio (\ref{ratio}) one does not have to perform any sophisticated calculations. 
We consider the scattering of $\sim 10$ MeV energy neutrinos, so that their wavelengths are much larger 
than the characteristic nuclear size of a few fm. Therefore, one can safely expand the nuclear matrix elements 
in series in $q$, or in  neutrino energy $E$, as $q$ is bounded by $E$. Here is how the inelastic matrix element of the 
$\mu=0$ component of the isoscalar vector current $J_\mu^{(0)}$
between the deuteron bound state 
and $np$ continuum will look like in this expansion:
\begin{eqnarray}
\label{expansion}
\langle d| \exp(i{\bf q r}^{(n)}) + \exp(i{\bf q r}^{(p)})  |np\rangle
\;\;\;\;\;\;\;\;\;\;\;\;\;\;\;\;\;\;\;\;\;\;\;\;\;\;\;\;\;\;\\
= 2\langle d| np\rangle + i {\bf q}\cdot \langle d| {\bf r}^{(n)} + {\bf r}^{(p)}  |np\rangle 
-\fr{q_kq_l}{2}\langle d| r_k^{(n)}r_l^{(n)} + r_k^{(p)}r_l^{(p)}|np\rangle
= -\fr{q_kq_l}{4}\langle d| r_kr_l|np\rangle,
\nonumber
\end{eqnarray}
where ${\bf r}^{(n)} ,~ {\bf r}^{(p)} $ are the position operators for the neutron and the proton. 
The zeroth and first order terms in $q = |{\bf q}|$ are trivially zero due to the orthogonality of 
the wave functions (${\bf r}^{(n)} + {\bf r}^{(p)} $ is the center-of-mass operator and cannot 
mediate inelastic transitions). In the last line we have introduced 
the relative position vector ${\bf r} = {\bf r}^{(n)} - {\bf r}^{(p)} $,
and the quadratic in $r$ operator can be further separated into the isotropic "charge-radius" 
and quadrupole components. For the $0^+ \to 2^+$ transition in  \ctw\ only the quadrupole part will matter. 
It is of course instructive to revisit the SM deuteron breakup \cite{Butler}, and observe that 
isoscalar vector component of the standard weak current gives a very minor contribution to the total 
cross section at low $E_\nu$ due to this $q^2$ suppression of the amplitude. The SM rate is of course dominated by the 
isovector axial-vector current that corresponds to the difference of nucleon spins ${\bf s}^{(n)} - {\bf s}^{(p)}$,
an operator that has non-zero inelastic matrix elements even in the $O(q^0)$ order.

We perform the calculation of the NCB-induced deuteron breakup using the "zero-radius" approximation 
of the initial and final state wave functions, 
\begin{eqnarray}
\psi_{\rm in}({\bf r}) = \fr{\sqrt{2\kappa}}{\sqrt{4\pi }r}\exp(-\kappa r); ~~ 
\psi_{\rm f}({\bf r}) = \exp(i{\bf pr})\\
\kappa = \sqrt{2E_b \mu} = 45 ~{\rm MeV} ; ~~ p^2={\bf p}^2 = 2\mu(E - E_f - E_b).
\end{eqnarray}
Here $E$ and $E_f$ are the initial and final energy of \nub, $E_b =2.2$ MeV is the absolute value of the deuteron binding energy,
and $\mu \simeq (m_n+m_p)/4$ is the reduced mass of the proton-neutron two-body system. Relative momentum $p$ 
of the final state is fully determined from the neutrino energies, as the recoil of the deuteron center-of-mass is 
negligible. Parameter $\kappa$ is the familiar "bound state momentum", related to the inverse size of the 
deuteron, $\kappa \sim R_d^{-1}$, and its relative smallness reflects large spatial extent of the deuteron. 
In a slightly excessive for the problem at hand language our calculations correspond to the leading order 
of the pionless effective field theory \cite{Butler,Pionless}. They can be systematically improved if needed, 
or treated with the more elaborate nuclear physics tools (see {\em e.g.} \cite{Kubodera}). None of this
will of course change the order of $q$ in which the effect first occurs. 

Straightforward calculations give the differential over the final neutrino energy cross section:
\begin{eqnarray}
\frac{d\sigma_{d\to np}}{dE_f}= \frac{G_B^2 E_f^2 m_p}{8\pi^2}
\fr{\kappa^5 p}{( p^2 + \kappa^2)^6}\left[  E^2E_f^2 + 
\fr{12 p^4}{5\kappa^4}(E^4-\fr{2}{3}E^3E_f +\fr{10}{9}E^2E_f^2 -\fr{2}{3}EE_f^3 +E_f^4)  \right]
\label{dtonp}
\end{eqnarray}
The result for $d\sigma_{d\to np}$ 
shows  a $O(E^4\kappa^{-4})$ suppression in agreement with $O(q^2)$ of the deuteron matrix element
and in agreement with (\ref{ratio}). Judging by the size of the subleading corrections in the SM calculations \cite{Butler}
we expect this answer to hold within $\sim 20\%$ accuracy. 

\begin{figure}
\centering
\includegraphics[width=0.85\textwidth]{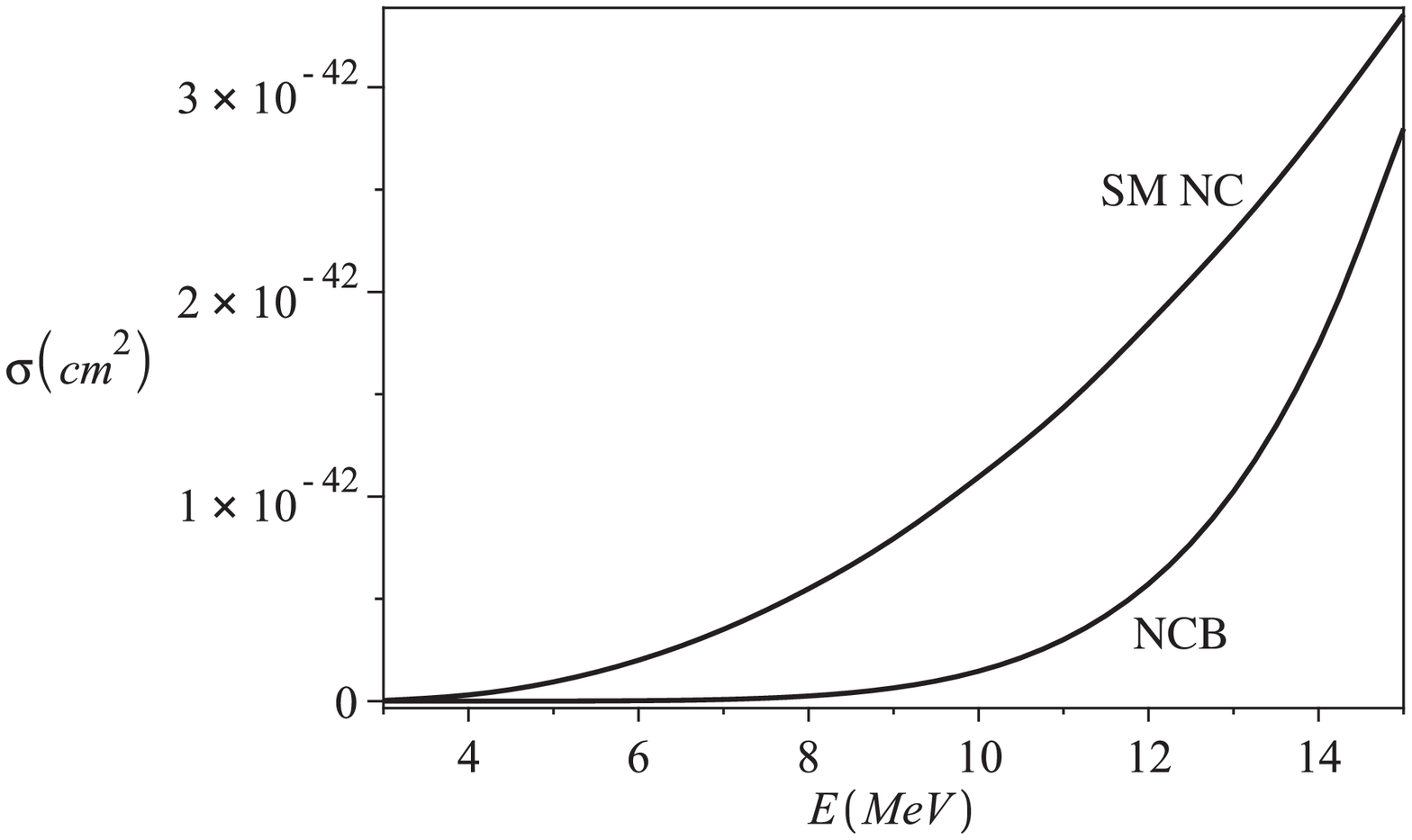} 

\vspace{0.5 cm}

\hspace{-0.5cm}\includegraphics[width=0.85\textwidth]{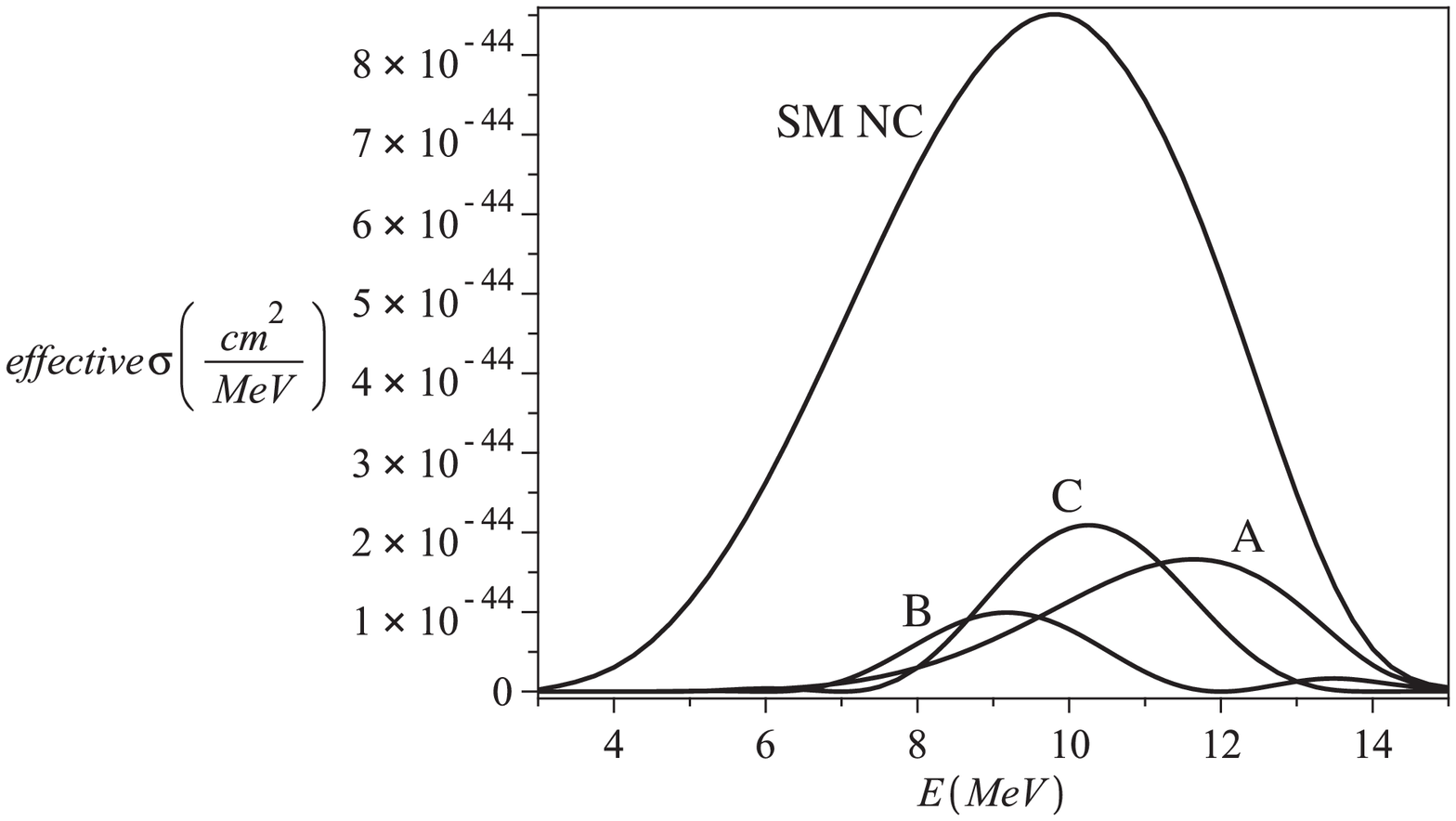}
\caption{Top pannel: Deuteron breakup cross section for the SM NC processes (top curve) and for the 
NCB \nub-neutrinos (bottom curve). The NCB cross section is plotted for ${\cal N}^2$ enhancement factor of $10^4$.  
Bottom pannel: same cross sections convoluted with \bore\ energy distribution and the energy dependent part of 
the oscillation probability. The top curve is the SM NC distribution of effective cross section, 
and A, B, C are the same for the NCB with $E_0=\infty, ~12, ~14$ 
MeV and $\neffm^2=10^4$. The areas under the curves give the proportion of neutrons produced via SM NC and NCB processes.}
\label{dnp}
\end{figure} 

The final integral over $E_f$ in the interval from $0$ to $E-E_b$ 
gives the total NCB deuteron breakup cross section. In Figure 2, upper pannel, we plot $\sigma_{d\to np}$
for the Standard Model neutrino and for \nub\ with the choice of enhancement factor ${\cal N}^2=10^4$. 
As expected, the NCB cross section has a faster rise with neutrino energy due to quadrupolar nature of the NCB interaction. 
In Figure 2, lower pannel,  we also show the convolution of the cross section with the energy 
distribution of boron neutrinos times the energy-dependent part of the oscillation probability, 
$2\sin^2(\pi E_0/E )$. As in the previous subsection, the NCB rates are considered 
for large $E_0$ and for $E_0 = 12,~14$ MeV, while the enhancement factor is kept at 
$\neffm =100$. The areas under curves give the total effective cross sections,
and when multiplied by $\Phi_{^8{\rm B}}$, correspond to the breakup rate per deuteron.
 For the three NCB cases considered here have the following comaprison to the SM rate:
\be
\fr{\sigma_{\rm NCB}}{\sigma_{\rm SM~NC}}\simeq \fr{\neffm^2}{10^4}\times (0.14,~0.06,~0.13)~~ ~{\rm at}~ ~E_0 = \infty;~ 12~{\rm MeV};~14~{\rm MeV}.
\ee
A 15\% increase in the neutron production rate at SNO can be tolerated, and if one chooses a 
sizeable $\theta_b$ so that the active neutrino flux is slightly less than SM+SSM predict, the 
total neutral current rate may not even change. 
We conclude that SNO NC events leave enough room for the possible $\neffm^2 \sim O(10^4)$ (and slightly 
higher) enhancement of the NCB rate.

We now address the $\ctwm\to\ctwm^*$ activation due to \nub. To avoid the complications 
arising from nuclear physics we shall assume that both the ground state and the first excited state 
of \ctw\ are given by $3\alpha$ configurations. This is a very well justified assumption, which leads to 
a relation between the matrix elements of the baryonic current and electric current,
\be
\langle 0^+ |  J_i^{(0)}  | 2^+ \rangle = 2 \langle 0^+ |  J_i^{\rm em}  | 2^+ \rangle = 
2 \fr{(E_{2^+} - E_{0^+}) q_j}{6} \langle 0^+ |  Q_{ij}^{\rm em} | 2^+ \rangle
\label{relation}
\ee
Only the lowest order in $q$ terms are retained here, and the 
$\mu=0$ component can be restored from gauge invariance. The factor of 2 in (\ref{relation})
comes from the fact that the baryonic charge of $\alpha$-particles is twice larger than its electric charge.
The information on the value of the transitional quadrupole moment $\langle Q_{ij}\rangle$ 
can be extracted from the $\ctwm^*$ decay width $\Gamma = 1.08\times 10^{-2}$ eV:
\be
\label{quadr}
\overline{|\langle  Q_{ij}^{\rm em}\rangle|^2}= \fr{90\Gamma}{\alpha (\Delta E)^5} = ( 3.3 ~{\rm fm} )^4,
\ee 
where the value of quadrupole moment squared is averaged over the 
arbitrary projection of the $2^+$ angular momentum, and $\Delta E = 
E_{2^+} - E_{0^+}=E-E_f = 4.439$ MeV. Note that we define 
electric quadrupole and electric current {\em without} the "$e$",
and account for $\alpha$ explicitly in (\ref{quadr}). The total 
inelastic cross section for the \nub-induced $\ctwm\to\ctwm ^*$ transition is given by
\be
\sigma_{^{12}{\rm C}\to ^{12}{\rm C}^*}=
\fr{8G_B^2E^5(E-\Delta E)\overline{|\langle  Q_{ij}^{\rm em}\rangle|^2}}{81\pi}
\left[ 1-3x+\fr{39}{8}x^2-\fr{19}{4}x^3+\fr{39}{16}x^4-\fr{9}{16}x^5\right],
\ee
where $x = \Delta E/E$. The benchmark value for this cross section at $E=8$ MeV and ${\cal N}=1$ 
is $2.5 \times 10^{-48}~{\rm cm}^2$. 

With this cross section the effective rate of injection of 4.4 MeV gamma quanta in pseudocumene
(scintillating material used by the Borexino experiment) is estimated to be 
\be
R(4.4 ~{\rm MeV}) \sim (0.05-0.15)\times \fr{\gamma~{\rm injections}}{\rm 100~tons\times day} \times \fr{\neffm^2}{10^4}.
\label{borex}
\ee
This is not a large rate by any measure, but it is nevertherless comparable to the counting rates 
in 3-5 MeV window from \bore\  ES processes and from the $^{208}$Tl background events \cite{BorexB}. 
The actual counting rate should be obtained by applying to (\ref{borex}) the efficiency factor that the collaboration 
can extract from their calibration data and simulations. At this point we can only conclude that 
there must be some sensitivity to NCB at $\neffm \sim 10^2$ level at the large scale neutrino detectors that 
use carbon-based scintillators. More definitive statements and perhaps stronger sensitivity to \nub\ can be 
derived from dedicated analyses. Moreover, the search for the 
extra $\gamma$-lines in a different energy range was already performed by the Borexino collaboration 
in connection with hypothetical Pauli-forbidden \ctw\ decays \cite{PauliB}. The search for NCB would represent a far 
less exotic physics cause in our opinion. One could also conduct similar searches of \nub-induced 
excitation of $^{16}$O nuclei using SNO and SuperK data.

\subsection*{4. Annual modulation of \nub\ rates}

In this section we would like to address the question of seasonal modulation of the NCB rate. 
The seasonal modulation of solar neutrino rate was observed by SNO and SuperK collaborations 
\cite{SuperK,Snoseason}. It exhibits full agreement with expected $\propto L^{-2}(t)$, 3.3\% modulation of the 
neutrino flux, with an appropriate minimum in the summer (Northern hemisphere). 
The hypothetical NCB elastic scattering rate will have the same modulation pattern as long as 
$\Delta m_b^2$ is large, or in other words  at  $E_0\gg E_{\nu~{\rm solar}}$.   
In the opposite limit of $E_0\ll E_{\nu~{\rm solar}}$ the modulation effects are suppressed because 
$\Phi_\nu \propto L_0^2L^{-2}(t)\sin^2[\pi E_0 L(t)(L_0 E)^{-1} ] \to (\pi E_0/E)^2$, which is time-independent.

However, it is easy to imagine that the flux of \nub\ neutrinos can have a more intricate seasonal modulation 
pattern. For example, if $E_0$ is between the maximum of the \bore\ neutrino spectrum and its end-point, 
the high-energy fraction of the distribution will have a higher flux in the summer. This is best illustrated 
in Figure 3, where the expected flux of \nub\ resulting from oscillations of boron neutrinos is 
convoluted with the time-dependent part of $P_b$,
$L_0^2L^{-2}(t)\sin^2[\pi E_0 L(t)(L_0 E)^{-1} ]$ at $E_0=12$ MeV. The two curves correspond to 
$t=t_{\rm perihelion}$ ($\sim$ 3 Jan) and $t_{\rm aphelion}$ ($\sim$ 4 Jul). Although on average there is more 
\nub\ neutrinos 
arriving to the Earth in January, in the most relevant range of energies, $E > 10 $ MeV, the flux in July is 
larger. Therefore, for this fraction of neutrinos there is a phase reversal, and the elastic scattering rates 
will reflect that. 

\begin{figure}
\centering
\includegraphics[width=1\textwidth]{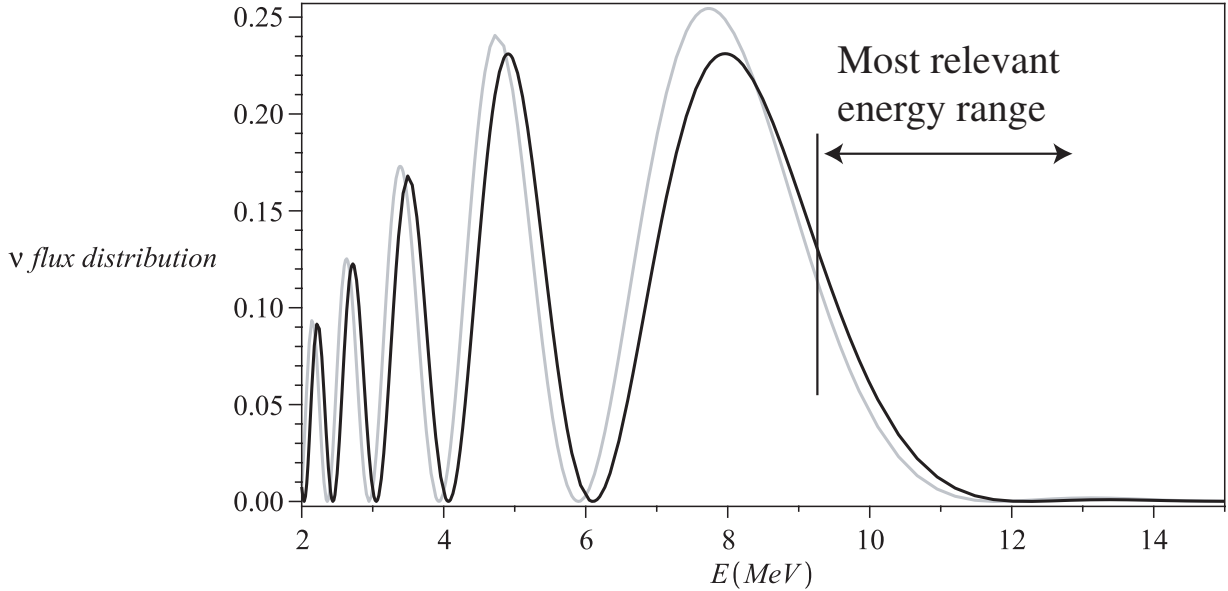} 
\caption{Boron \nub\ neutrino flux modified by the time-dependent part of the oscillation probability 
$ 2L_0^2L^{-2}(t)\sin^2[\pi E_0 L(t)(L_0 E)^{-1} ]$ with $E_0=12$ MeV. The black curve is for July, and 
the gray curve is for January. Although the total integral under the gray curve is bigger than under the black one, it is the 
high-end of the spectrum that would determine rates at the existing DM detectors, where July 
rates are larger. }
\label{Seasons}
\end{figure} 

Next we calculate the expected seasonal modulation in the counting rate, 
\be
\label{sinus}
\fr{dR_{\rm mod}}{dE_r} = \fr12\left( \left.\fr{dR}{dE_r}\right|_{\rm Jul} -\left.\fr{dR}{dE_r}\right|_{\rm Jan}    \right),
\ee
for NaI detectors using the quenching factor from Eq. (\ref{quenching}).
We would like to remark in passing that for some ranges of neutrino energies 
there can be a significant departure from a simple time-sinusoidal function, 
but to observe such effects one would probably require very high statistics and very good energy
resolution. Modulation rates,  $dR_{\rm mod}/dE_r$, as defined in Eq. (\ref{sinus}) are plotted in Figure 4
for the same three choices of $E_0$ and $\neffm=100$ as before. One can see that indeed modulation of both signs is 
possible, and that the rate of modulated NCB signal at this \neff\ is 
indeed probed by DAMA/LIBRA experiment \cite{DAMA}, which is sensitive to modulation amplitudes $O(10^{-2})$cpd/kg/keVee. 

\begin{figure}
\centering
\includegraphics[width=0.90\textwidth]{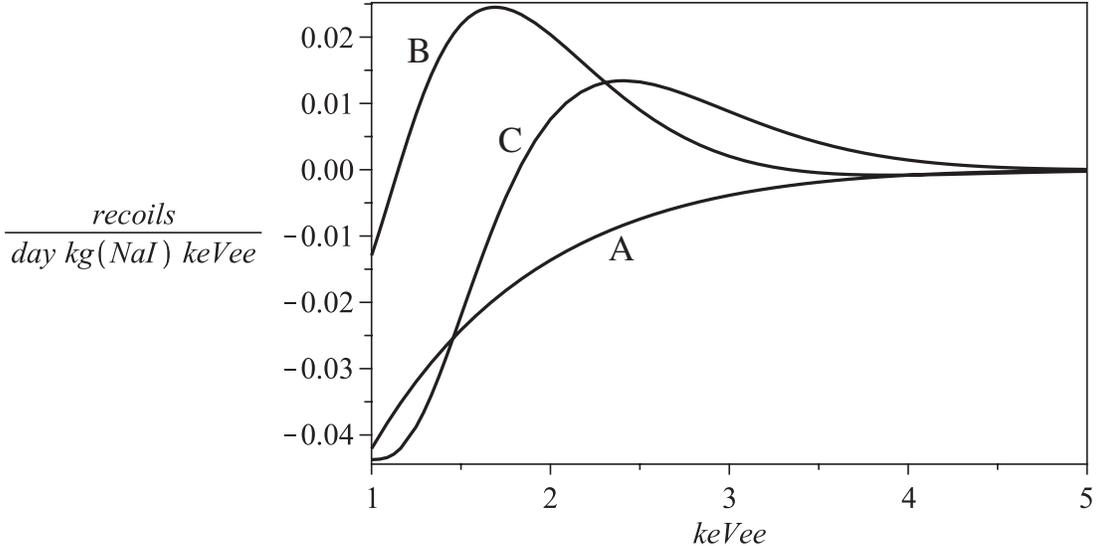} 
\caption{Modulation of the counting rate in recoils/kg(NaI)/keVee for \nub-scattering on 
Na. As before, A curve is for large $E_0$, B is for $E_0 = 12$ MeV and 
C is for $E_0=14$ MeV while $\neffm^2=10^4$. Both signs of modulation are possible.  }
\label{Modulation}
\end{figure}

Is it possible that \nub-Na elastic scattering is behind the DAMA/LIBRA seasonal modulation anomaly? 
The magnitude of the predicted modulation can be in a very good agreement with DAMA results \cite{DAMA}.
Moerover, as we saw in the previous section, $\neffm^2\sim 10^4$ is thus far consistent with other observations 
and constraints (and with simultaneous explanation of 
\co\ low $E_r$ anomaly). Of course the phase of the DAMA results will require $E_0$ to be in the right range.
But even then, would the early July maximum be consistent with DAMA/LIBRA claims of the oscillation phase? 
The best fit point for the maximum is about 4-5 weeks different from the $t_{\rm aphelion}$ \cite{DAMA}. 
It would be interesting to know if the early July maximum is actually excluded by DAMA data,
and the criticism expressed in Ref. \cite{phase} about the errors on the phase being too tight be
properly addressed by the collaboration. 

Going away from \nub\ idea, one can also notice that many other exotic physics explanations of DAMA 
signal can be invoked (if it allows to tolerate $\sim$ 1 month phase shift). 
For example, the emission of solar axions with their subsequent absorption in DM detection experiments 
can be a cause of low-energy ionization signal \cite{aSun}. On this picture, one can super-impose the oscillations of axions 
into some "sterile axions" with the oscillation length similar to $L_0$ in order to break the monotonic
$L(t)$-dependence, and flip the phase of the modulation, achieving results similar to those of Figure 3 and 4. 
%Also, as pointed out in \cite{phase}, 
%seasonal modulation of cosmic rays at Earth's surface may produce additional effects. Given a 

\subsection*{5. Discussion and conclusions}

The oscillation of SM neutrinos to some new neutrino state on the way from the Sun to the Earth 
is a realistic possibility.
We have shown that there exists the whole class of models where neutrino physics 
beyond SM can be probed by the low-threshold DM detectors, which become equally sensitive 
or even more sensitive to this type of neutrinos than the large scale neutrino detectors. 
Such models require that new neutrino states \nub\ (or modified SM model neutrino interactions 
in the spirit of Eq. (\ref{howdoyoulikeit}))
couple almost exclusively to the baryon current. The isoscalar vector properties of this current 
lead to a very strong enhancement of the elastic over inelastic scattering, $\sigma_{\rm elastic}/\sigma_{\rm inelastic}
\sim 10^8$, providing an unexpected competitiveness factor to small scale experiments such as \co. 
We have shown the effective strength of the NCB can be much larger 
than the weak-scale value without being in conflict with any of the observational data. 

We have also shown that the recent anomalies in direct DM detection, such as DAMA and \co,
can be explained by the $\nu_{\rm SM}\to \nu_b$ oscillation of \bore\ neutrinos 
with subsequent scattering of \nub\ on Ge and Na nuclei. (This statement 
relies on the assumption of $t_{\rm aphelion}$ being consistent with DAMA/LIBRA modulation phase.)
This may look counter-intuitive at first, but we have shown that the phase flip of seasonal modulations 
is possible for the high-energy end of the \bore\ spectrum. This is a very speculative 
explanation (and perhaps equally speculative as the WIMP recoil explanation), but it
is interesting enough to motivate further studies. In particular, we believe that the 
Borexino collaboration can perform the search of the \nub-activated 4.4 MeV line in \ctw,
and probably surpass the sensitivity to the NCB enhancement factor $\neffm$ of $100$. 
At the same time it seems apparent that further technological developments of 
low-threshold WIMP/\nub\ detectors are required. Should the current low-energy anomalies in DM detectors
firm up to constitute a definitive signal of new-physics-induced recoil, some significant efforts and different mass targets 
would be required to observationally distinguish between the low-mass WIMP and \nub\ signals. 

Below, we would like to discuss further implications of the models involving new neutrino states with 
enhanced baryonic currents. 
\begin{itemize}

\item {\em Collider implications.} If the $G_B \ga 100 G_F$, and if the new 
interaction is truly contact, the proton-antiproton collisons
will lead to strong new sources of missing energy signals in $\nu_b \bar\nu_b$ channel 
and will most likely be excluded by the Tevatron experiments. This will not happen, however, in models 
of relatively weakly coupled mediators with sub-GeV mass. Therefore the collider searches should be able
to place an {\em upper} bound on $m_V$. 

\item {\em Fixed target implications.} A GeV-scale \ub\ baryonic vector, the carrier of 
NCB interaction, can be produced in the 
collisions of energetic proton beams with a target. Immediate decays of these vectors 
will generate a flux of \nub\ state that can be searched for at near detectors via their 
NCB interactions. This is very similar to the ideas of the "MeV-scale DM beams" 
discussed previously in \cite{target}. There can be also implications for the 
terrestrial anti-neutrino physics, as matter effects induced by $V$-exchange can be large 
for \nub\ antineutrinos \cite{Nelson}. Enhanced neutral currents of \nub\ neutrinos may 
help explaining the long-standing puzzle of the LSND anomaly \cite{LSND}, perhaps borrowing some elements of the 
recent suggestion \cite{Gnin}. It also has to be said that over the last two years a lot of efforts have been 
invested in systematically searching for the "kinetic mixing" (or hypercharge) portal (see {\em e.g.}  \cite{ourguys} 
and references therein).
Barionic portal is another example of a perfectly safe from the model-building perspective way of 
introducing stronger-than-weak forces at low energy, and therefore it should be systematically 
searched for using proton-on-target facilities. But perhaps the most NCB-search effective type of experiments 
to perform with proton beams is the proposal \cite{CLEAR} of a neutrino-nucleus elastic scattering detector. 

\item  {\em Cosmological implications.} A new light neutrino state (and two neutrino states if right-handed 
copy of \nub\ is also light) can be at borderline of what is allowed by early cosmology and observations 
of light elemental abundances (most recent analysis can be found in \cite{bbn}). Is the model 
with much enhanced baryonic currents has a chance to be consistent with these constraints? Actually, despite the interaction 
strength of 100 or 1000 of $G_F$, \nub\ will decouple from thermal plasma {\em earlier} than the SM neutrinos. 
That is because its thermalization rate will be proprtional to the baryon-to-photon ratio, which is a
small number $O(10^{-10})$. Therefore actual decoupling of \nub\ may happen with the decays and annihilations 
of abundant hadronic species at temperatures of $\sim 100$ MeV, and therefore BBN bounds 
from over-population of radiative degrees of freedom can be easily evaded. 
 
\item  {\em Astrophysical implications.} Another interesting aspect of \nub\ models is their NCB production 
in stars. In the SN  \nub\ will not provide new effective energy sinks because they would not escape 
freely the explosion zone. However one should expect that a comparable to the SM number of \nub\ neutrinos is 
created, so that one could detect them  using the same DM/\nub\ detectors. Should a nearby SN explosion 
happen, the existing neutrino scintillator detectors can pick up the \nub-NCB signal that would appear as a much enhanced 
$\nu_\mu,\nu_\tau$-NC signal considered in \cite{Beacomtoday} (modulo the uncertainty in effective temperature for 
\nub). 

\item  {\em Rare decay implications.} Relatively large NCB currents should open new channels 
for the missing energy decays of $B$ and $K$ mesons. As argued in this paper, the conservation of 
the baryonic current makes it a relatively safe portal compared to {\em e.g.} scalar or axial-vector portal. 
Nevertheless, if the $K\to \pi V$ decay is kinematically allowed, it may lead to the underlying 
two-body signature of $K\to \pi$ plus missing energy decays, making it an appealing target for the 
next generation of precision kaon physics experiments. 

\end{itemize} 

\subsection*{Acknowledgements}

I would like to thank B. Batell, D. McKeen, J. Pradler and I. Yavin for  helpful discussions.
I am also grateful to A. de Gouvea, R. Harnik and J. Kopp for pointing out inconsistencies 
of the neutrino oscillation picture in version 1 of this work. 
The work is supported in part by the Government of Canada through NSERC and by the Province of Ontario through MEDT.

\end{document}